\documentclass[review]{elsarticle}
\usepackage{hyperref}
\usepackage{graphicx,amsmath,amssymb,epstopdf,array,booktabs,makecell}
\pdfoutput=1

\journal{Elsevier Journal of Computer Communications}

\begin{document}

\begin{frontmatter}

\title{Random Forest Resource Allocation for 5G Systems: Performance and Robustness Study}


\author[mymainaddress]{Sahar Imtiaz\corref{mycorrespondingauthor}}
\cortext[mycorrespondingauthor]{Corresponding author}
\ead{sahari@kth.se}
\author[mymainaddress]{Hadi Ghauch}
\author[mythirdaddress]{Muhammad Mahboob Ur Rahman}
\author[mysecondaryaddress]{George Koudouridis}
\author[mymainaddress]{James Gross}

\address[mymainaddress]{KTH Royal Institute of Technology, School of Electrical Engineering, Stockholm, Sweden}
\address[mythirdaddress]{Information Technology University (ITU), Lahore, Pakistan}
\address[mysecondaryaddress]{Radio Network Technology Research, Huawei Technologies, Kista, Sweden}

\begin{abstract}
Next generation cellular networks will have to leverage a further cell 
densification to accomplish the ambitious goals with respect to 
aggregate multi-user sum rates.
It is well known that this requires much more coordination among the 
transmission points in the system to balance interference with terminal 
performance.
Traditionally, this has been limited by the coordination capabilities in 
the backbone, nevertheless with the cloud radio access network (CRAN) 
architecture these limitations are likely to be overcome.
This shifts the attention back to applicable resource allocation, which 
need to be applicable for very short radio frames, large and dense sets 
of radio heads, and large user populations in the coordination area.
So far, mainly channel state information (CSI)-based resource allocation schemes have been proposed for this task.
However, they come at a considerable complexity while also incurring a 
significant price in terms of CSI acquisition overhead on the system.
In this paper, we study an alternative approach which promises lower 
complexity while also having a lower overhead.
In particular, we propose to base the resource allocation in 
multi-antenna CRAN systems on the position information of user terminals 
only.
Based on the user positions, we further propose the application of 
Random Forests as supervised machine learning approach to determine the 
multi-user resource allocations.
This likely leads to lower overhead costs, as the acquisition of 
position information requires less radio resources in comparison to the 
acquisition of instantaneous CSI.
In addition, once a corresponding data structure is learned, the 
complexity for determining a resource allocation for a given user is low 
as well.
After presenting our design, we extensively benchmark it with the 
following findings: (I) In general, learning-based RA schemes can 
achieve comparable spectral efficiency to CSI-based scheme; (II) If 
taking the system overhead into account, learning-based RA scheme 
utilizing position information outperform legacy CSI-based scheme by up 
to 100\% ; (III) Despite their dependency on the training data, Random 
Forests based RA scheme is robust against position inaccuracies and 
changes in the propagation scenario; (IV) The most important factor 
influencing the performance of learning-based RA scheme is the antenna 
orientation, for which we present three approaches that restore most of 
the original performance when facing random antenna orientations of the 
user terminal.
To the best of our knowledge, these insights are new and indicate a 
novel as well as promising approach to master the complexity in future 
cellular networks.
\end{abstract}


\end{frontmatter}

\section{Introduction}
Presently, developing the fifth generation (5G) cellular networking technology is one of the key research topics in the academic and industrial community. 
One of the key drivers for this is to ensure high data rate provision to all users, irrespective of their location and time of network access. 
Typically, 5G systems are attributed with the following service requirements:
\begin{itemize}
\item a 1000$\times$ increase in system capacity compared to Long Term Evolution-Advanced (LTE-A) systems \cite{5G_NetworkCapacity}.
\item at least 10$\times$ reduced end-to-end latency compared to LTE-A systems \cite{Petteri_IEEE_Access}, i.e. a round trip time (RTT) of less than 1 ms.
\item almost 100$\times$ increased energy efficiency in Joules/bit \cite{Petteri_IEEE_Access}.
\item support for medium to high mobility users, with high throughput and always-on connectivity requirements \cite{Imtiaz:2016:LRA}.
\end{itemize}

There is no doubt that in order to massively increase the system capacity, network densification is necessary, which directly leads to an increased interference in the system. 
The existing LTE system architecture does not allow to handle the required coordination to be able to cater such severe interference scenarios resulting from the network densification. 
Thus, new network architectures had to be devised, from which the cloud radio access network (CRAN) architecture \cite{CRAN_ELMAR2013} is a promising way forward for implementing such dense networks at relatively moderate costs, and hence, is a favoured architecture for 5G systems' deployment. 
In CRAN, the radio access units, formed from distributed antenna systems, are separated from the central processing units, that handle all the baseband processing. 
The central processing units, essentially being small cloud-like data processing units, are also connected to each other through a backbone allowing for fast coordination among the central processors. 
A single unit of the distributed antenna systems is called remote radio head (RRH), which when densely placed with other RRHs in an area of interest forms then an antenna domain (AN), and this set up is formally known as ultra-dense network (UDN) deployment~\cite{ngmn20155g}.
Such UDN deployments are ideal for achieving tight interference coordination between RRHs, leading to very high system capacity, and thus can achieve the aforementioned targets for 5G communication systems.

However, with very low envisioned frame times, while at the same time handling large amount of users in the antenna domain, the overhead for channel state acquisition in such UDNs is known to become excessive as the densification increases. 
This is in particular true for moderate to high speeds of the user terminals, where, as a consequence, the channel states fluctuate intensely~\cite{Shen_massiveMIMO}. 
Nevertheless, the densification in tendency also leads to more and more connections essentially being line-of-sight (LOS), leading in principle to lower statistical channel variability.
This motivates the consideration of alternative resource allocation approaches, which are not based on CSI acquisition, but utilize pure position information of the users in the antenna domain.
While such an approach potentially leads to a lower overhead, as position estimation based on beaconing requires a much lower overhead than CSI acquisition, it is open how to perform the resource allocation based on the positions of the users, which complexity this includes, and how such schemes perform in terms of key network performance indicators when benchmarked with CSI-based schemes. 
In order to overcome these challenges, we propose in this work the usage of machine learning for the resource allocation (RA) based on position information of the users.
This promises significantly lower complexity in comparison to CSI-based schemes, while it is open which spectral efficiency it can achieve and which robustness such a learning-based approach has to the changes in the propagation scenario or inaccuracies with respect to the provided position information. 
In order to address these issues, we resort in particular to Random Forests, as straightforward general machine learning algorithm and data structure, that nevertheless is well known for its inherent robustness \cite{Breiman_RF}, \cite{IEEE_2013_Oscar}.

Related work with respect to machine learning, resource allocation and CRAN is still quite sparse: 
In \cite{RAAC_2016_GC}, the authors propose a resource allocation scheme based on linearisation of Mixed Integer Non-Linear Program (MINLP) for mobile users present in 5G systems with CRAN architecture, where they formulate the problem as maximization of network throughput, with a constraint on maximum network capacity. 
The authors in \cite{D2D_CSI_comp} have shown that resource allocation based on CSI is much expensive in terms of system overhead compared to location-based resource allocation scheme in the context of device-to-device (D2D) communications, which are an integral part of 5G systems' design. In such case, when perfect CSI is used for resource allocation, the system overhead can be as large as about 25\% of the system capacity.
The authors in \cite{Joao_WCNC_2012} use the reinforcement learning, a machine learning technique, for adaptive modulation and coding in orthogonal frequency division multiplex-multiple input, multiple output (OFDM-MIMO) based 5G systems. In our previous work \citep{Imtiaz:2016:LRA}, we used the random forests algorithm as a binary classifier for allocating resources to users present in CRAN-based 5G system. In that case, the random forests classifier was coupled with a system scheduler, which validated the prediction provided by the random forest, and then the appropriate resources were allocated to serve the given set of users in the system. Though we evaluated the robustness of the resource allocation scheme based on the binary random forest classifier for different system parametrization, but the scope of such investigations was quite limited.
We thus conclude that all of our above addressed challenges are still open, some of which we will investigate in this work.

The main contributions of our paper are as follows:
\begin{itemize}
\item We present a design of a learning-based RA scheme for 5G systems by using Random Forests as multi-class classifier, which essentially predicts the modulation and coding scheme to be used for a given position of a terminal. 
\item Through numerical evaluation we demonstrate the basic efficiency of the approach: While in typical deployment the learning-based RA scheme can achieve a comparable spectral efficiency to CSI-based schemes, if the corresponding overhead is considered as well, the learning-based approach outperforms CSI-based approaches significantly.
\item We demonstrate the robustness of the proposed scheme with respect to different variations of users' position accuracy, showing that even for quite large variations the learning-based approach can still provide good performance.
\item As most important parameter, we study the impact of a random orientation of the user antennas. If this is not accounted for, this leads to a strong performance decrease in case of the learning-based approach, while we discuss several compensation schemes, which overcome this challenge.
\end{itemize} 

The remaining paper is structured in the following manner: Section`\ref{Sys_model} presents the system model and the detailed problem statement. Some background information on machine learning and Random Forests algorithm is presented in Section~\ref{LbRA}, along with the details for the design of learning-based RA scheme. The performance evaluation of the proposed scheme is then elaborated in Section~\ref{Results}. Section~\ref{C&F} concludes the paper finally, accompanied by the discussion of the future work.


\section{System Model}
\label{Sys_model}

\begin{figure}[!]
\centering
\includegraphics[width=5cm]{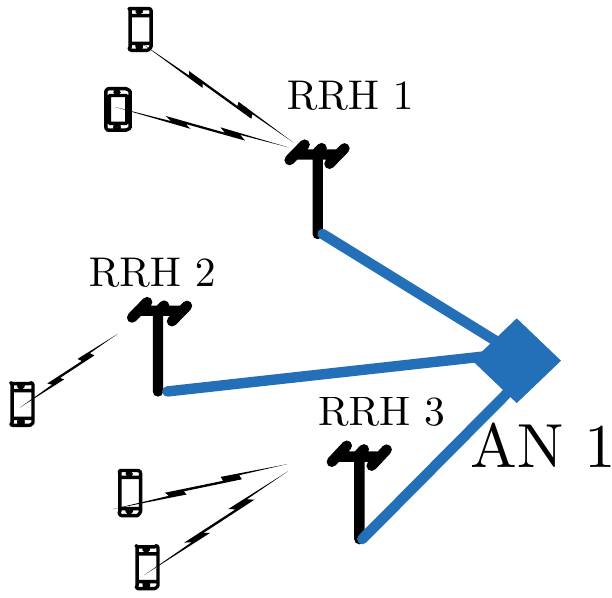}
\caption{The CRAN architecture for 5G system}
\label{CRAN}
\end{figure}

Consider the CRAN system as shown strongly simplified in Figure~\ref{CRAN}. 
Over a given area of the CRAN system, there are $N$ users which are served by $R$ remote radio heads (RRHs). 
The RRHs are connected by a fast back-haul to an aggregation node (AN), which performs all baseband processing as well as provides the gateway to the deeper backbone.
The CRAN operates in time division duplex (TDD) with time slots (referred to in the following as transmission time intervals - TTIs) of duration $T_{\mathrm{f}}$ while in the following we only consider the downlink communication direction.
The operating frequency of the system is $f_{\mathrm{c}}$, with a system bandwidth $W$. 
The RRHs are considered to be densely deployed in the consider CRAN (an example could be placing distributed antenna systems on top of street lights \cite{Petteri_IEEE_Access}).
Users are roaming freely within the CRAN with varying velocities and direction. 
Each RRH and user is equipped with $N_{\text{Tx}}$ and $N_{\text{Rx}}$ antennas, respectively. 


\subsection{Resource Allocation and Channel Model}
For a given downlink frame, each RRH serves at least one user in the system.
The assignment of users to RRHs is performed at the AN, which is the RA unit of the CRAN and therefore also determines all other resource parameters to let users be served by RRHs. 
Apart from the user assignment, in this work we focus primarily on the determination of the transmit beams per RRH, the receive filters per user, as well as the selection of appropriate modulation and coding schemes (MCSs). 
At each RRH there exists (similar) fixed set of transmit beams $B_{\text{Tx}}$ that the AN can choose from in order to serve the assigned users. 
Similarly, for each assigned user there exists a fixed set of receive filters $B_{\text{Rx}}$ from which the AN can choose one.
We also assume a fixed set of modulation and coding schemes to exist from which the AN selects one for each RRH/user assignment.
User assignments and resource allocation vary from frame to frame in general.
For the determination of the user assignments and the resource allocations, the AN can utilize different information. 
On the one hand, we assume that it can acquire channel state information. 
On the other hand, also the position information of the users can be obtained by the AN.
Acquisition of both comes at a certain price, which we discuss below in more detail.

In the following, we restrict ourselves to the case where each RRH is serving only one user in a given time slot.
Due to the density of the RRH deployment, this nevertheless can lead to significant interference between the RRHs if the resource allocation is not performed carefully, i.e. in general we assume an interference-limited propagation scenario.
In general, the downlink channel between each RRH $r$ and user $n$ is characterized by spatial parameters (like angle-of-arrival (AoA), and angle-of-departure (AoD), as well as scatterers' density), frequency-related parameters (operational frequency of the system and the Doppler shift) and time-related parameters (such as power delay profile and change in users' position). 
Given a certain allocation of users to RRHs, transmit beams and receive filters, the signal-to-interference-and-noise ratio (SINR) of a user $n$ allocated to a certain RRH for a given time $t$ is given by:
\begin{align}
\label{SINR_equation}
\gamma_{n, t}(\phi_n^a, \phi_n^d) = \frac{P_{n, t}(\phi_n^a, \phi_n^d)}{ \sigma^2 + \sum\limits_{ \substack{m=1 \\ m \neq n } }^{N} P_{m, t}(\phi_n^a, \phi_m^d) },
\end{align}

\noindent
where, $P_{n, t}$ is the received signal power for user $n$ at time $t$ and is given by:
\begin{align}
\label{received_power_equation}
P_{n, t}(\phi_n^a, \phi_n^d) = & \ P_{\text{Tx}} h_{\text{PL}}^2 \cdot \ | \pmb{U}(\phi_n^a)^\dagger \pmb{H}_{n, t}(\phi_n^a, \phi_n^d) \pmb{V}( \phi_n^d )|^2 .
\end{align}
Here, $P_{\text{Tx}}$ denotes the transmit power allocated per RRH, $h_{\text{PL}}^2$ denotes the pathloss, $\phi_n^a$ is the azimuth AoA of user $n$, and $\phi_n^d$ is its azimuth AoD. 
$\pmb{U}(\phi_n^a)$ is the receive filter with the main beam focused in the direction closest to $\phi_n^a$, and $\pmb{V}(\phi_n^d)$ is the transmit beamformer with the main beam located in the direction closest to $\phi_n^d$. 
$\pmb{H}_{n, t}(\phi_n^a, \phi_n^d)$ is the channel matrix for an instance of time $t$ for a given $\phi_n^a$ and $\phi_n^d$, and $\sigma^2$ is the noise power. 
$(.)^\dagger$ denotes the Hermitian of a vector or a matrix.
Throughout the time frame we assume this SINR to remain constant in time and frequency.


Given the SINR per slot, the system utilizes a certain set of modulation and coding schemes to convey the backlogged information to the corresponding user.
This transmission is nevertheless subject to block errors, which is captured through an appropriate link-to-system interface. 
We assume in the following an LTE-like link-to-system interface, where the channel SINR of a user is modelled to a channel quality indicator (CQI) level, implying the usage of a corresponding MCS with certain spectral efficiency and leading to a specific packet error rate. 
Thus, the choice of the MCS leads to a certain payload size that can be sent over the channel. 
In the following, we consider the modulation and coding schemes with bandwidth efficiencies for the different CQI levels for the link-to-system interface given in~\cite{afifi2012radio}.

\subsection{Overhead Modelling}
In order to acquire either the channel state information (in form of a complete characterization of the complex channel gains from each RRH to each user) or the positions of the users, the system needs to spend some overhead.
We model this overhead by considering a fraction of the TTIs consumed for acquiring the system state. 
More precisely, we consider the duration of a TTI to consist of a certain set $T_{\text{sym,total}}$ of symbols and $ f_{\text{sc,total}}$  OFDM sub-carriers. 
The position information of the users present in the system is acquired using so called narrow-band pilots, spanning the first symbol of the TTI but only requiring a few OFDM sub-carriers. Thus, using a single symbol, a multitude of user positions can be determined through narrow-band beaconing.
On the other hand, in order to acquire the CSI of the users, the so called full-band pilots need to be employed. These span potentially multiple symbols and require the entire system bandwidth. 
The adjacent CSI-sensing pilots are scheduled based on the cyclic-prefix compensation distance, as explained in \cite{Petteri_IEEE_Access}, to avoid inter-carrier interference. 

Based on these parameters, the overhead for position acquisition per TTI can be calculated as:
\begin{align}
\label{OH_pos}
OH_{pos} = \frac{T_{\text{sym},pos} \times f_{\text{sc},pos} }{T_{\text{sym,total}} \times f_{\text{sc,total}}}.
\end{align}
Here, $T_{\text{sym},pos}$ is the number of OFDM symbols used for position estimation of users in the system, and $f_{\text{sc},pos}$ denotes the number of sub-carriers used in the positioning beacon. 

Similarly, for CSI acquisition per TTI, the overhead can be computed as:
\begin{align}
\label{OH_CSI}
OH_{CSI} = \frac{T_{\text{sym},CSI} \times f_{\text{sc},CSI} }{T_{\text{sym,total}} \times f_{\text{sc,total}}},
\end{align}
\noindent
where $T_{\text{sym},CSI}$ and $f_{\text{sc},CSI}$ denote the number of OFDM symbols and the number of sub-carriers, used for CSI acquisition of users present in the system in a TTI, respectively. 
As an example, if one system TTI is 1 ms, then it will comprise of 5 TDD frames. 
For the case when very few users are present in the system, a single positioning beacon spanning one time symbol and the whole range of frequency sub-carriers will be considered for computing the positioning-based system overhead. 
For CSI-sensing in such scenario, a number of full-band time symbols will be used, depending on the number of users present in the system. 
With more users being a part of the system ($\sim$25), the overhead for the CSI-sensing and position acquisition increases, however, in general it increases much stronger in case of the CSI acquisition.

\subsection{Problem Statement}
A central question for future 5G systems based on the CRAN architecture relates to the efficient resource allocation.
Traditionally, for multi-cell, multi-antenna deployments, CSI-based resource allocation schemes have been proposed, operating either in a centralized or distributed fashion.
While CRAN alleviates the bottleneck of the coordination overhead between the transmission points (in this case, the RRHs), CSI acquisition nevertheless incurs significant overhead with respect to the reduction of the time fraction usable for payload transmission per frame.
This is in particular true for large user populations that have to be served within one coordination area, where the user speeds are moderate to high.
In addition, it is well known that CSI-based resource allocation schemes selecting user assignments, transmit beams and receive filters lead to a significant computational complexity.
To mitigate this overhead and complexity, in general we are interested in this paper in alternative approaches.
For illustration purposes, we consider in the following the maximization of the sum-rate as objective function for the resource allocation per TTI. 

In particular, we are interested in approaches that rely solely on the position information of the users, as this is in general much easier to acquire in terms of overhead.
This opens up the question how the resource allocation is to be performed per TTI. 
For this, our approach is to leverage machine learning, and couple a learned data structure to the scheduler at the AN of the CRAN.
This is considerably less complex in comparison to CSI-based resource allocation schemes, but leads to other potential issues, mainly related to the robustness of the learned structure to changes in the live environment, as well as to dependencies between the amount of learned training data versus the resulting system performance in general.
We are thus interested in (a) a design for such a learning-based resource allocation based on user positions; (b) a performance evaluation of CSI-based resource allocation in comparison to the learning-based approach; (c) a study on the robustness of the learning-based approach in case of different parameter variations.
In the following, we will first discuss a learning-based approach which utilizes the Random Forests algorithm, and afterwards present a deep performance evaluation.

\section{Learning-based Resource Allocation Scheme}
\label{LbRA}
In this section, we initially recap machine learning and in particular the Random Forests algorithm. This is followed by the introduction and discussion of our design of a learning-based RA scheme.

\subsection{Machine Learning and Random Forests Algorithm}
Machine learning is a tool used for making a computer program or a machine ``develop new knowledge or skill from the existing and non-existing data samples to optimize some performance criteria" \cite{alpaydin2014introduction}. 
The machine learning algorithms are of three types: \emph{supervised} learning algorithms, where the output variable(s) to be predicted by the algorithm is(are) known beforehand; \emph{unsupervised} learning algorithms, where the output variable(s) [or label(s)] are not known in advance for the training data samples; and \emph{reinforcement} learning algorithms, where the learning algorithm receives feedback from the environment itself to reinforce the learning pattern for the target. 

\begin{figure}
\centering
\includegraphics[width=10cm]{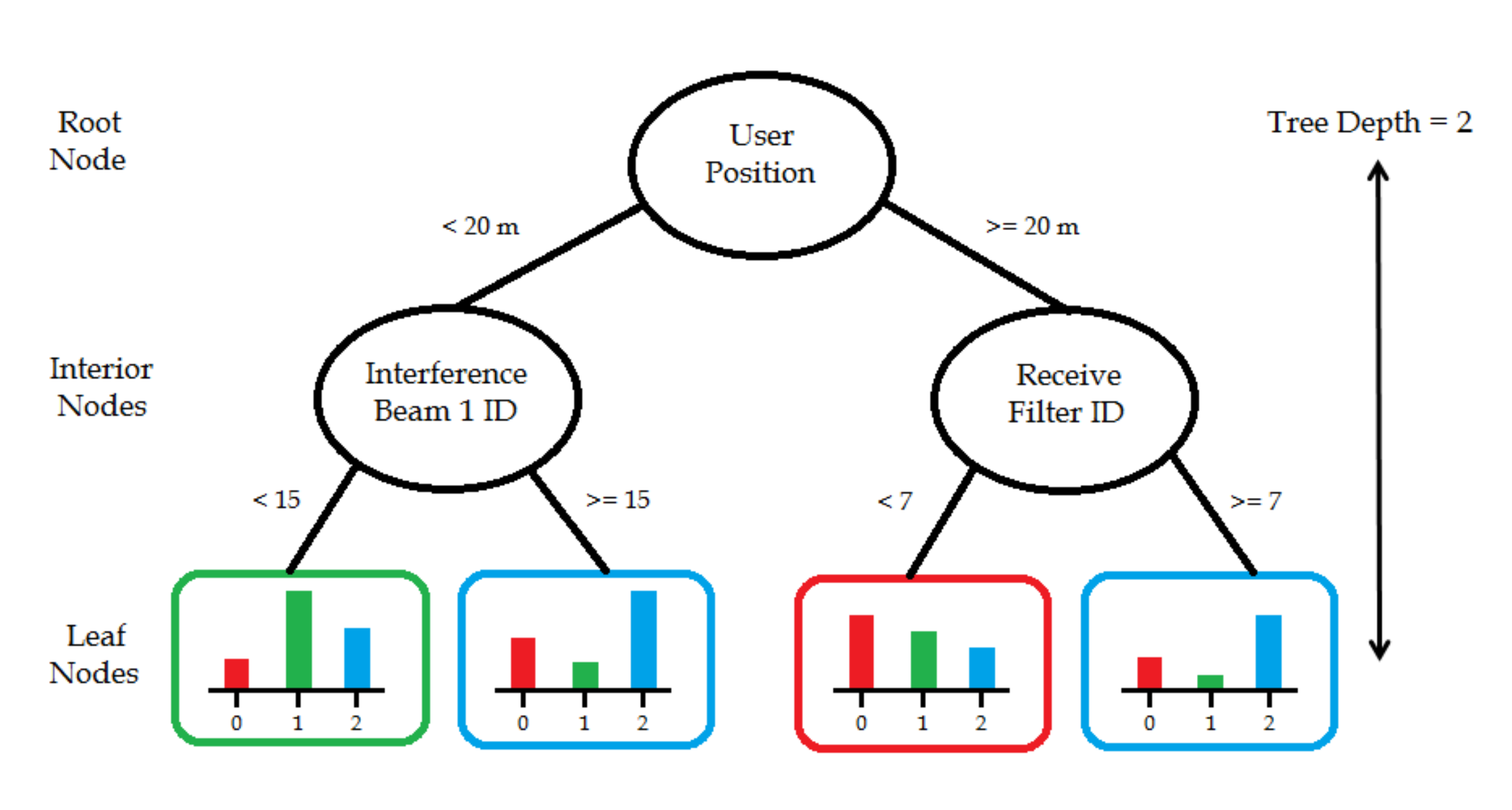}
\caption{An example of a binary random decision tree}
\label{random_tree}
\end{figure}

In this work, we employ the Random Forests algorithm~\cite{Breiman_RF}, which is a \emph{supervised} machine learning technique, for designing the learning-based RA scheme. 
As the name suggests, the algorithm uses a combination of multiple random binary decision trees, which make up the \emph{forest}, for predicting one or a set  of outcome(s).
For this, in general a set of training data needs to be collected which is provided to the Random Forests algorithm to generate the decision trees.  
A prerequisite for this is that the training dataset $\pmb{X}$ is split into two parts: a set of data characteristics or features $\pmb{F}$, and a set of output variables $\pmb{Y}$. 
Each instance $\pmb{x}_i$ of the training dataset is called an \emph{input feature vector}. 
The algorithm then constructs $T_n$ binary random trees, each with a depth $T_d$, using the different features (selected randomly) in the training dataset. 
Each tree typically consists of a root node, one or more interior nodes and terminates at leaf nodes, as shown in the sample tree in figure \ref{random_tree}. 
The leaf nodes store the output variable(s), technically called a `vote', and the output variable predicted by the algorithm is the mode of those votes from all trees in the random forest. 
Once a forest has been trained (during the operational phase), the input features of a new (and potentially unknown) instance is presented to the decision forest, leading to a prediction (through the voting of the trees) of the output variable.
More detailed descriptions about the working of the random forests algorithm can be found in \cite{Breiman_RF}.
In general, Random Forests are known to be easy-to-use but robust machine learning data structures, which motivated our choice in this study.

\subsection{Random Forest Learning-based RA Scheme}
Our principle design for a learning-based RA scheme relates to the Random Forest providing decision on the resource allocation per TTI.
Very different approaches can be applied here, however, we strive in this work for a simple but robust design that operates on a user base. 
That means, the output relates to the prediction of a resource allocation variable for a specific, dedicated user while a set of input variables is provided.
Specifically, as input we consider first of all the acquired user position estimate $\mathcal{P}_n$ of the dedicated user. 
In addition, the transmit beam $B_{\text{Tx}}$ used for serving the dedicated user as well as the received filter $B_{\text{Rx}}$ is provided. 
Finally, the interfering beams from the neighbouring RRHs affecting the transmission reception for the dedicated user are provided.  
Using all these parameters, we construct the input feature vector for building up the random forest. 
The output variable $y$ is the MCS parameter, which is tied to different CQI levels, and thus the Random Forest is used for \emph{multi-class classification}, where $y$ has different CQI levels as labels, or classes. 
Thus, the random forest algorithm essentially works as a scheduler in the system, where it is used for predicting the CQI level for a given set of input features for a dedicated user $n$. 
Based on the output from the random forest, the appropriate modulation and coding scheme can then be determined. 
Note in particular, that the input feature vector contains all interfering beams in the system, as well as preselected beam and receive filter for the dedicated user.

An important aspect relates to the transmit beam and receive filter to be applied in the construction of the input feature vector.
For this, a preprocessing is applied where for an essential set of gridded positions in the area of the considered antenna domain the optimal transmit beams and receive filters for combinations of user positions are determined according to the considered objective function, i.e. the sum-rate of the scheduled users.
Given a new, and potentially not considered positioning of the users, in the first step this is matched to the closest set of known gridded position combinations for which a set of transmit beams and receive filters are known. 
This set of transmit beams is then used for generating the input feature vector together with the receive filter of the dedicated user currently considered.
The input feature vector is then forwarded to Random Forest, where each data sample is parsed through the random trees in the forest data structure to get a prediction for CQI level (and thus for the modulation and coding scheme) as the output variable.

Due to the inherent property of the random forests algorithm, this learning-based RA scheme can be expected to be somewhat robust to noisy data.
Therefore, in theory, having inaccuracy in position estimates of the users present in the system should not have much impact on the overall system performance. 
Nevertheless, in general this robustness only holds up to a certain limit.
In the performance evaluation section further below, an important aspect is to determine these limits (and potentially discuss remedies). 
 
 
\subsection{Dimensioning of the Random Forest for Resource Allocations} 
Before a forest-based data structure can be applied, it first needs to be learned off-line based on training data.
For this, a considerable amount of instances needs to be collected, for instance from an optimal CSI-based resource allocation scheme.
The training of the random forest optimizes then the forest data structure for accuracy. 
Here, an important aspect relates to the dimensioning of the forest itself, as it impacts the training and test accuracy. 
Dimensioning relates to the depth of the trees as well as the number of trees to be used in the forest.

The training accuracy is obtained by using a subset of training data for validation of the constructed random forest model. 
Once an sufficiently dimensioned random forest structure has been found, a test data set is then used to compute the test accuracy of the model by passing each instance of the test data set through each of the random trees in the model. 
The higher the number of correctly predicted output by the model (whether for the validation data set, or the test data set), the higher will be the accuracy. 
However, having a very high training accuracy is not an indicator of an appropriate random forest structure. 
It could be the case that the random forests structure works perfectly for the training data set, but shows a low accuracy for test data set. 
Such a random forests structure is then an over-fit to the training data. 
For building a robust random forest structure, we need to vary the number of trees $T_n$ in the forest, as well as the depth of the trees $T_d$, in such a way that the model achieves a fairly high training accuracy but shows good test accuracy for any test data set with similar input feature vector composition. 
Hence, for some data collected from a first system set-up (see the evaluation section for details) we study in Table~\ref{table_RF} the training and test accuracy obtained for different parametrization of the random forests structure. 
Based on these investigations, we used the best possible random forests model for the design of the learning-based RA scheme, with 100 trees, each with a depth of 10.

\begin{table}
\caption{Training and test accuracy for different parametrization of random forests model}
\label{table_RF}
\centering
\begin{tabular}{|c|c|c|c|}
\hline
$T_n$ & $T_d$ & Training Accuracy (\%) & Test Accuracy (\%) \\
\hline
5 & 3 & 89.7 & 90.5\\
\hline
10 & 3 & 89.75 & 90.25\\
10 & 5 & 94 & 92.75\\
10 & 10 & 99.5 & 93\\
\hline
50 & 5 & 94.75 & 92.75\\
50 & 10 & 99 & 93.5\\
\hline
100 & 5 & 94 & 92.75\\
\textbf{100} & \textbf{10} & \textbf{99.5} & \textbf{93.25}\\
\hline
200 & 10 & 99.5 & 93.25\\
\hline
300 & 10 & 99.5 & 93.25\\
\hline
\end{tabular}
\end{table} 

Finally, it is to be noted here that the random forest does not suffer severely from a bias problem \cite{Imtiaz:2016:LRA}, since almost each class in the output variable is sufficiently represented in the input feature vectors used for training the random forest. 
Once the random forest achieves an optimal training accuracy, based on different parametrization of the algorithm, it is available for predicting the output variable for the test dataset generated at run-time of the considered CRAN system.

\section{Performance Evaluation}
\label{Results}
In this section, we will first present our evaluation methodology. 
Then, we present various results on the performance of the learning-based RA scheme in comparison to benchmark schemes. 
We finally consider various aspects on the robustness and performance limitation of the proposed learning-based RA scheme. 

\begin{figure}[!]
\centering
\includegraphics[width=7cm]{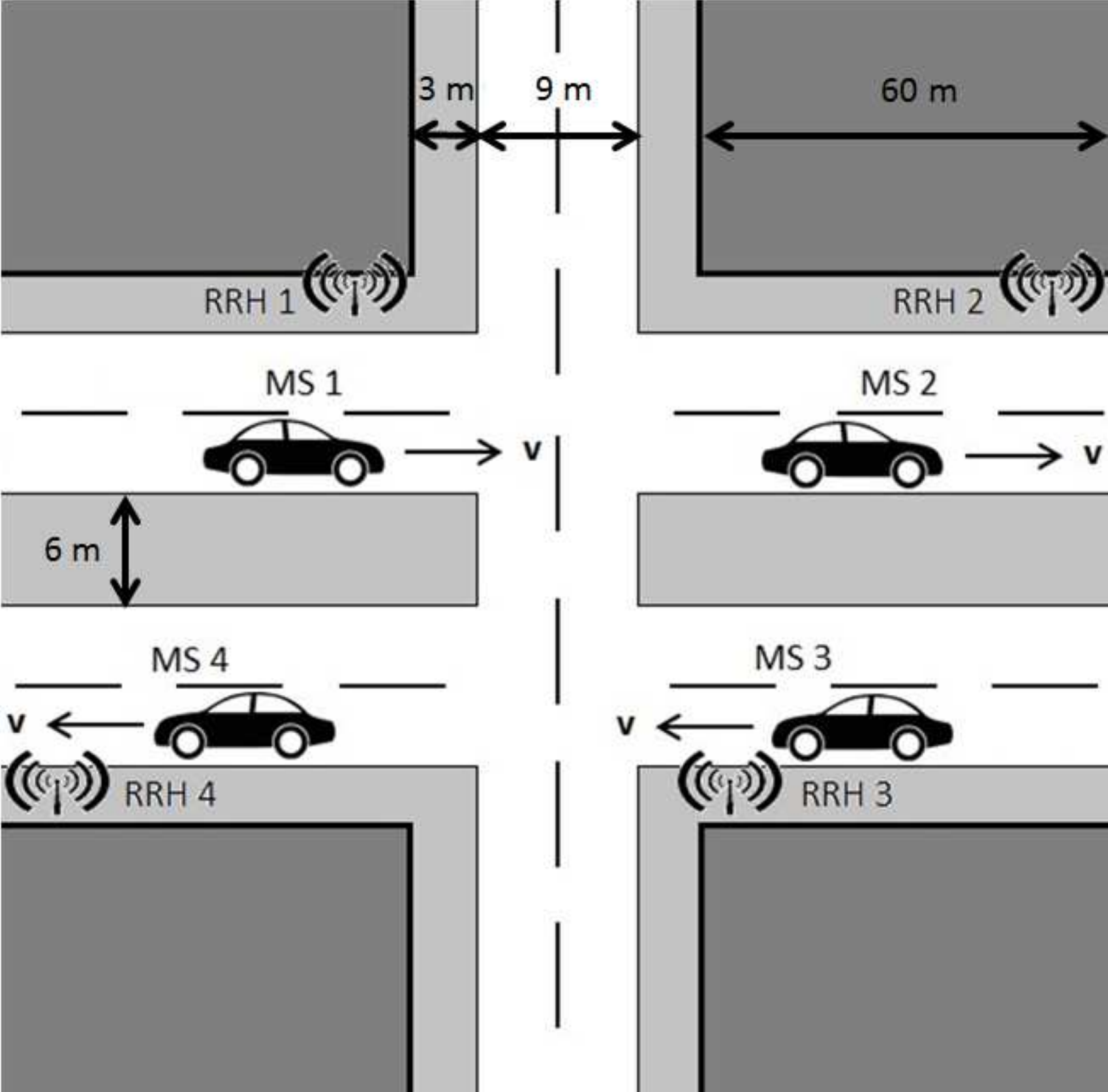}
\caption{The simulation scenario; each RRH serves one user}
\label{sim_scenario}
\end{figure}

\subsection{Evaluation Methodology}
The performance evaluation of the proposed learning-based RA scheme is done by performing simulations using the discrete event simulator \emph{Horizon}~\cite{ref_29_D1_SofA}.
Figure~\ref{sim_scenario} shows the simulation scenario, comprising 4 RRHs, each serving a single user. 
This represents a simpler multi-RRH, multi-user scenario for 5G CRAN system where only inter-RRH interference exists. 
A fixed set of transmit beams is designed using geometric beamforming, with an angular separation of 3$^{\circ}$. 
The receive filters are designed in the same way as the transmit beams, but the angular separation is kept as 12$^{\circ}$. 
Other parameter settings for the simulation set up are given in Table~\ref{table_sim}. 
Since downlink communication is assumed in the system model, therefore, channel coefficients for TDD-based downlink are extracted for each RRH-user link in the simulation scenario, using the map-based METIS channel model for Madrid grid \cite{METIS_D1.2}. 
A ray-tracer based channel model was implemented for this purpose, the details for which can be found in~\cite{Location_BF_UDN}. 
Note that in Table~\ref{table_sim} $h_{\text{Tx}}$ refers to the height of the RRH antennas from the ground, while $h_{\text{Rx}}$ refers to the user antenna height from the ground. 
\begin{table}
\caption{Parameter Settings}
\label{table_sim}
\centering
\begin{tabular}{c c}
\hline
Parameter & Value\\
\hline
$f_c$ & 3.5 GHz\\
$BW$ & 200 MHz\\
$R_{\text{Tx}}$ & 8\\
$N_{\text{Rx}}$ & 2\\
$h_{\text{Tx}}$ & 10 m\\
$h_{\text{Rx}}$ & 1.5 m\\
$P_{\text{Tx}}$ & 1 mW\\
$TTI$ & 0.2 ms\\
$v_{Rx}$ & 30 m/s\\
\hline
\end{tabular}
\end{table}

Depending on the investigation scenario, as mentioned in Section~\ref{Sys_model}, the training dataset is constructed as the next step, using the procedure outlined in Section~\ref{LbRA}. 
These training datasets are used to construct the multi-class random forest model, where we rely on the implementation provided by OpenCV~\cite{opencv_library}. 
As for the parametrization of random forests algorithm, we set the number of trees $T_n$ = 100, each with a depth $T_d =$ 10, with the number of randomly selected features for each node split set as 3 (which is closest to the most balanced setting of $\sqrt{I}$ according to~\cite{Breiman_RF}). 
A total of 100 user positions, per user, are selected randomly from a set of 1000 user positions generated by Horizon, per user, to create training datasets of 0.25 million samples for each investigation scenario. The output from the random forests model is used to compute the user goodput, for each time instance $t$, using the following formula:

\begin{align}
\label{goodput_eq}
Goodput_{n, t} = \frac{(1 - BLER_{PS}) \times PS_{n, t}}{TTI}.
\end{align}
Here, $BLER$ is the block error rate for the packet size $PS$ assigned to the user $n$ at time $t$, and $TTI$ denotes the transmission time interval of the system. 
The system goodput is computed by taking the sum of the user goodput for each time instance, and its average over all considered 100 user positions is used for performance evaluation.

\subsection{Evaluation Results for the Proposed Learning-based RA Scheme}
We initially start with benchmarking the raw goodput for different schemes based on perfect system status knowledge, i.e. position or CSI.
In detail we consider the following schemes:
\begin{itemize}
\item The proposed learning-based RA scheme; where the multi-class random forest is used for allocating appropriate resources.
\item A random packet size allocation scheme; this uses the same input features as used for the learning-based scheme, but assigns a randomly selected packet size to serve a given user. This scheme serves as a benchmark to directly determine the value of learning the modulation and coding scheme for a given input feature vector.
\item A geometric-based RA scheme; where the user position information is used for allocating the transmit beams and receive filters for serving a given user, while again selecting the modulation and coding scheme randomly. This scheme benchmarks, in addition, the value of the pre-processing.
\item A legacy CSI-based scheme; for simplicity we consider here a scheme that determines the optimal transmit beam and receive filters based on the given CSI. This serves as an upper bound on the system performance.
\end{itemize}
Note that in the following investigation we do not consider the impact from the overhead model.

Table~\ref{results_GBRA} presents the average system goodput for all the above mentioned comparison schemes. 
We initially recognize that the learning-based RA scheme achieves a performance quite close to the CSI-based scheme. 
In contrast, the scheme based on random packet assignment performs a lot worse than the learning-based RA scheme, thus signifying the importance of learning the correlation between different system parameters. 
The geometric-based RA scheme shows the lowest system goodput compared to the goodput obtained from the CSI-based scheme; the reason being a severely interference-limited system considered for the given case. 
Also, since the selection of packet size for serving a given user with known position is done at random, therefore, the system goodput degrades even further. 
From this point onwards, we will provide a comparison of results for the proposed learning-based scheme with the CSI-based RA scheme only, since the random packet allocation scheme as well as the geometric-based RA scheme reap off very low system goodput.

\begin{table}
\caption{Comparison of system goodput (in \%age) for different schemes w.r.t. CSI-based RA scheme}
\label{results_GBRA}
\centering
\begin{tabular}{c c c c}
\hline
\thead{Learning-based\\ RA Scheme} & \thead{Random Packet Size\\Allocation Scheme} & \thead{Geometric-based\\ RA Scheme} & \thead{CSI-based\\ RA Scheme}\\
\hline
95.6\% & 3.45\% & 2.32\% & 100\%\\
\hline
\end{tabular}
\end{table}

We next turn to the the evaluation of the different approaches taking the system overhead into account.
Since we set the system TTI duration to 1 ms, we assume 5 TDD-frames to be used for position, or CSI, acquisition and data transmission for all users present in the system. 
This serves as basic parametrization for the overhead calculations presented in equations \ref{OH_pos} and \ref{OH_CSI}.
Our goal is to study the impact of the overhead on the performance of the learning-based and CSI-based RA schemes as the number of users in the system (for which the state information needs to be collected) grows.
Note that we consider at this step still all state information to be perfectly accurate (i.e. the position information as well as the CSI).
\begin{figure}[!]
\centering
\includegraphics[width=10cm,height=6cm]{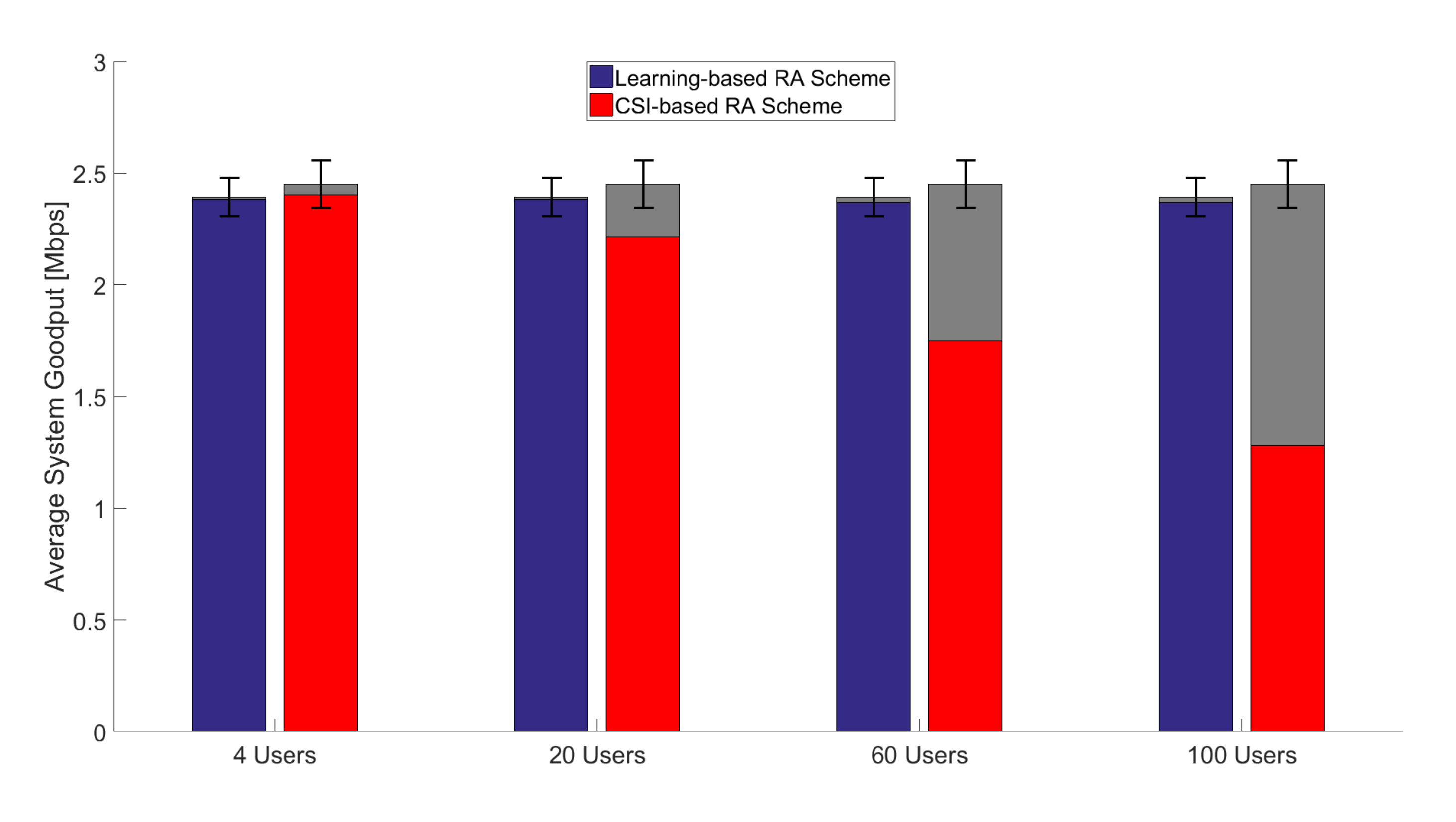}
\caption{Effect of overhead on average system goodput for different RA schemes, for perfect position estimates of all users}
\label{results_perfect_pos}
\end{figure}

Figure~\ref{results_perfect_pos} shows the results for average system goodput obtained using accurate user position information at all RRHs for the learning-based and CSI-based RA schemes. 
The colored bars show the effective average system goodput, i.e. the system goodput obtained after taking into account the effect of system overhead due to position beaconing or CSI sensing, while the underlying gray bars represent the system performance without taking the overhead into account. 
Overall, the proposed RA scheme achieves about 96\% of the system goodput achieved by the CSI-based scheme, without considering any overhead. However, if the system overhead is accounted for, we observe that the proposed scheme is either at par or better in performance compared to the CSI-based scheme for all possible number of users present in the system. 
In particular, as we increase the number of users in the system, the number of narrow-band beacons for acquiring users' position estimates increases per TTI, and thus the overhead scales up only marginally for the learning-based RA scheme. 
In contrast, the overhead for the CSI-based scheme grows much stronger with the increase in the number of users present in the system reaching up to 48\% of the frame time, showing that effective system performance degrades severely if CSI-based scheme is used for resource allocation in a system with high user density.

These two initial results are quite striking: Firstly, with respect to pure spectral efficiency, a learning-based RA scheme using position information only can achieve quite a good performance already in comparison to a CSI-based scheme. 
This holds at least for the considered system scenario, which nevertheless has been designed carefully and contains a typical level of detail for a system-level simulation of a 5G network.
Second, if the overhead or the state acquisition is factored in, due to the high cost of the CSI acquisition, the learning-based RA scheme can significantly outperform CSI-based approaches (up to 100 \% performance improvement).

\subsection{Robustness of Learning-Based RA Scheme}
\begin{figure}[!]
\centering
\includegraphics[width=10cm,height=6cm]{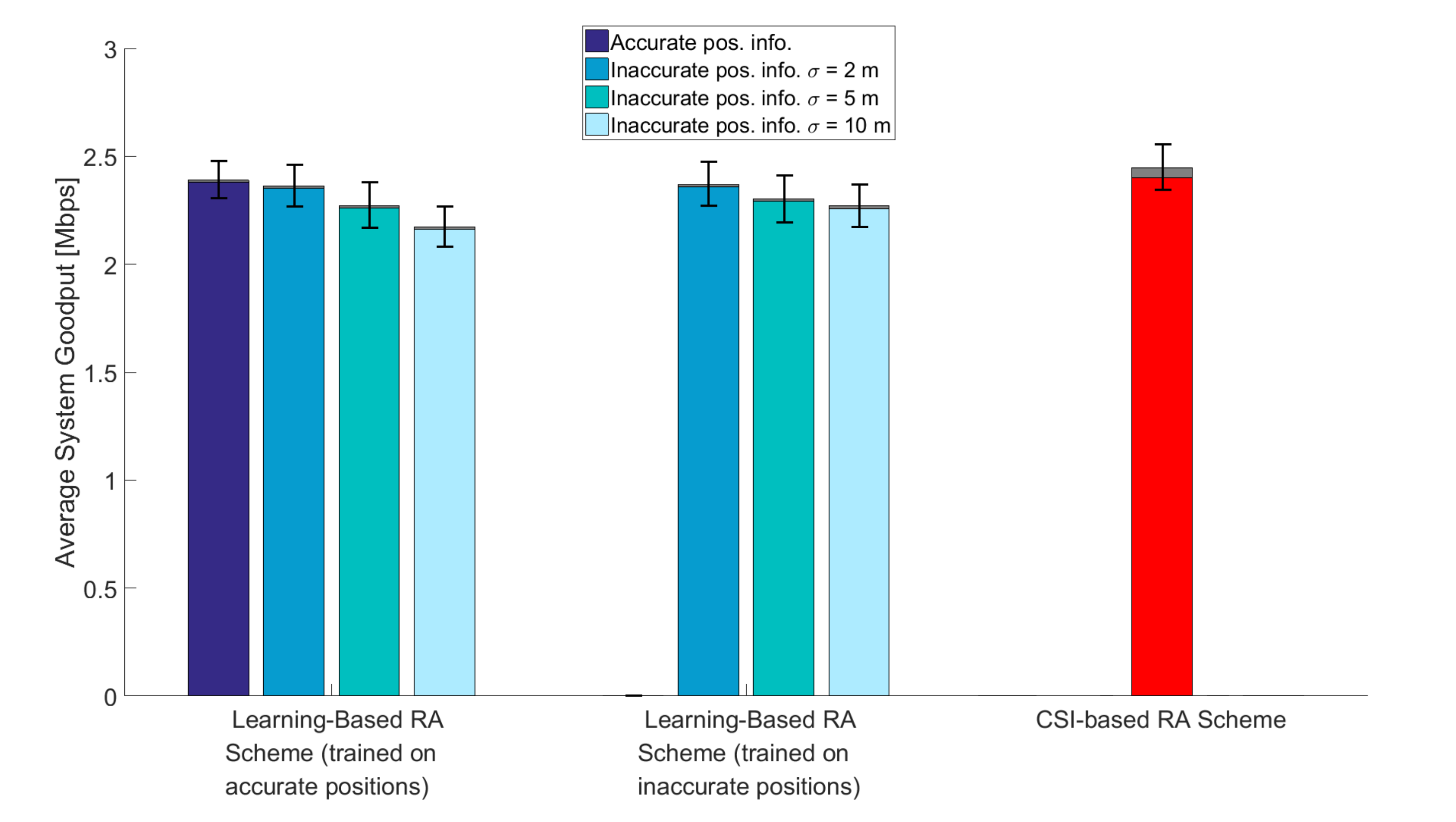}
\caption{Average system goodput for different schemes and various possibilities of available position information}
\label{results_all_pos}
\end{figure}

This performance advantage motivates a more thorough study on the robustness of our learning-based RA scheme.
We start with considering the most obvious potential source of inaccuracy influencing the learning-based scheme, namely the accuracy of the position information.
Figure~\ref{results_all_pos} shows the results for the average system goodput obtained when a random error is involved in the position estimation for the users being served by RRHs. 
It can be seen that the classifier trained on perfect user position information is enough to guarantee good system performance upto a certain degree of error involved in the position estimation. 
However, if the error margin in the user position estimates exceeds 2 m, the learning-based RA scheme trained on perfect user position estimates fails to provide satisfactory system performance. 
Better system goodput can be obtained by using the learning-based RA scheme trained on inaccurate position estimates, but the traditional CSI-based provides still about 10\% better effective system performance. 
This shows the robustness of the proposed scheme for small degrees of error involved in acquired user position information, but when the error margin becomes excessively large, the CSI-based RA scheme provides better effective system performance, when the best-case user density scenario is considered. 

\begin{figure}[!]
\centering
\includegraphics[width=10cm,height=6cm]{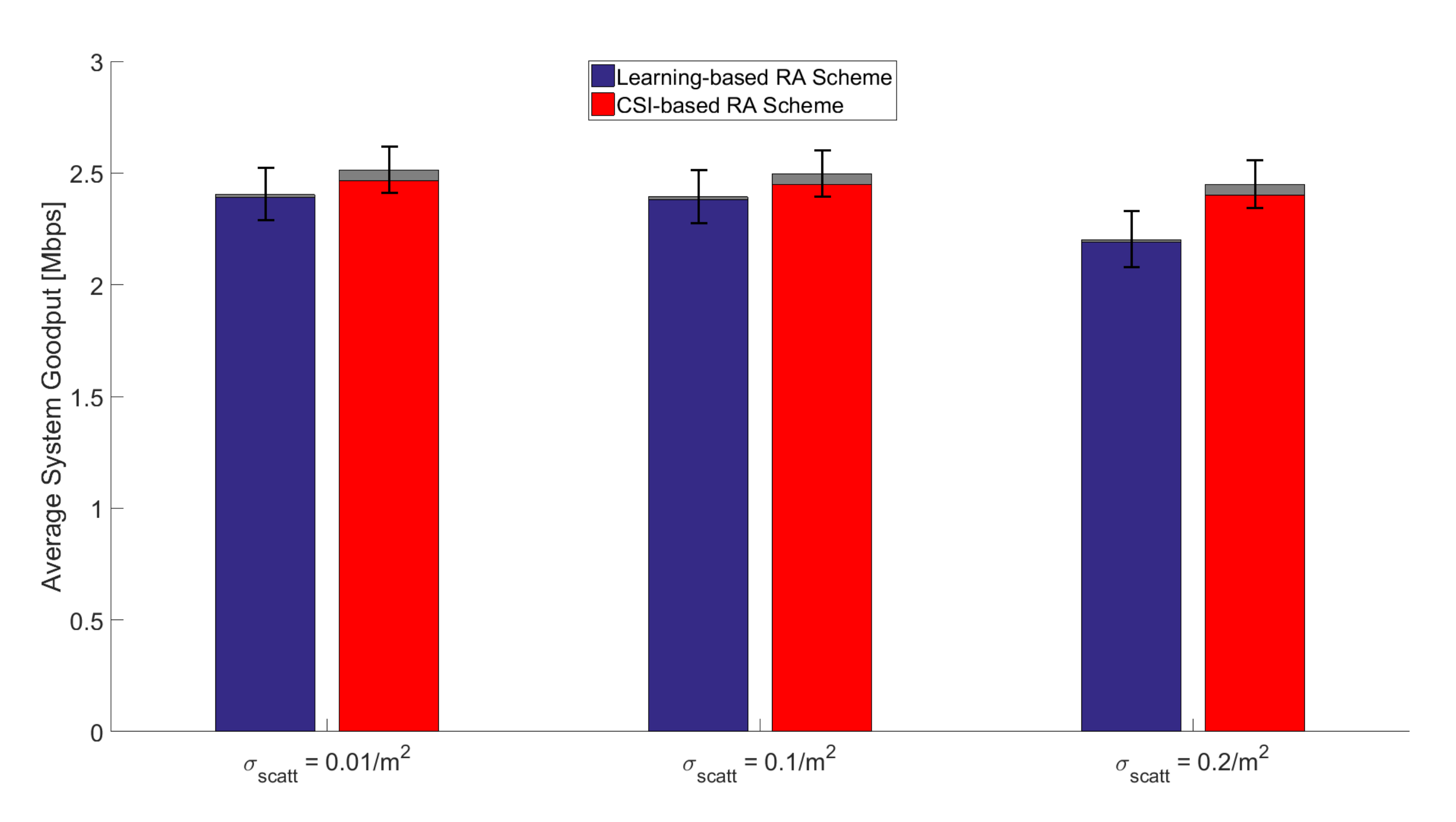}
\caption{Average system goodput for different scattering densities for perfect users' position information}
\label{results_all_scatt}
\end{figure}

We next turn to the question how sensitive the learning-based RA scheme is to a change in the propagation scenario in contrast to the one from which the training data has been acquired. 
Figure~\ref{results_all_scatt} shows the results for changing obstacle/scatterer density when the random forests model is trained only for a fixed system parameterization. 
We observe that the average system goodput varies only marginally with varying scatterers' density in the AD. 
Overall, the proposed scheme experiences only 7\% loss compared to the traditional CSI-based RA scheme in terms of effective system goodput obtained for all considered scenarios.

%

\subsection{Sensitivity to Random Antenna Orientation}
In the learning-based RA scheme, one of the allocated resources includes the receive filter, which is based on the beamforming in the direction closest to the direction of the received signal. 
For this to work perfectly, it is necessary to have the knowledge of the user equipment's (UE's) antenna orientation at the RRH serving the related user. 
The antenna orientation of the UE defines the radiation pattern of the receiving antenna, which dictates the selection of the receive filter. 
However, the UE antenna orientation is typically random and can therefore be defined in the local coordinate system (LCS), whereas the UE antenna orientation known at RRH is defined in the global coordinate system (GCS). In order to compute the correct direction of receive filter, the following transformation between GCS and LCS has to be used (based on the discussion given in the METIS Channel Model documentation \cite{METIS_D1.2}):

\begin{gather}
F_{GCS} (\theta, \phi) = \begin{bmatrix}
cos \varphi && -sin \varphi \\
sin \varphi && cos \varphi
\end{bmatrix}
\begin{bmatrix}
F_{\theta, LCS} (\theta ' , \phi ') \\
F_{\phi, LCS} (\theta ' , \phi ') \\
\end{bmatrix} .
\end{gather}

Here, $F_{GCS}$ represents the antenna radiation pattern in GCS, $\theta '$ and $\phi '$ are the elevation and azimuth angles in LCS, and $F_{\theta, LCS}$ and $F_{\phi, LCS}$ denote the radiation patterns of UE antenna in elevation and azimuth planes, respectively. $cos \varphi$ and $sin \varphi$ are the system transformation variables, given by \cite{METIS_D1.2}

\begin{align}
\label{cos_psi}
cos \varphi = \pmb{e}_{\theta, GCS} (\theta, \phi)^T \pmb{R}\:  \pmb{e}_{\theta, LCS} (\theta ', \phi ') ,\\
sin \varphi = \pmb{e}_{\phi, GCS} (\theta, \phi)^T \pmb{R}\:  \pmb{e}_{\theta, LCS} (\theta ', \phi ') .
\end{align}

\noindent
where, $\pmb{e}_{\theta, GCS}$ and $\pmb{e}_{\phi, GCS}$ are the basis vectors in GCS for elevation and azimuth planes, respectively, $\pmb{R}$ is the rotation matrix applied for correcting the angular orientation in GCS based on the orientation in LCS, and $\pmb{e}_{\theta, LCS}$ and $\pmb{e}_{\phi, LCS}$ are the basis vectors in LCS for elevation and azimuth planes, respectively. The derivation of the rotation matrix is given in \ref{App1}.

\begin{figure}[!]
\centering
\includegraphics[width=10cm,height=6cm]{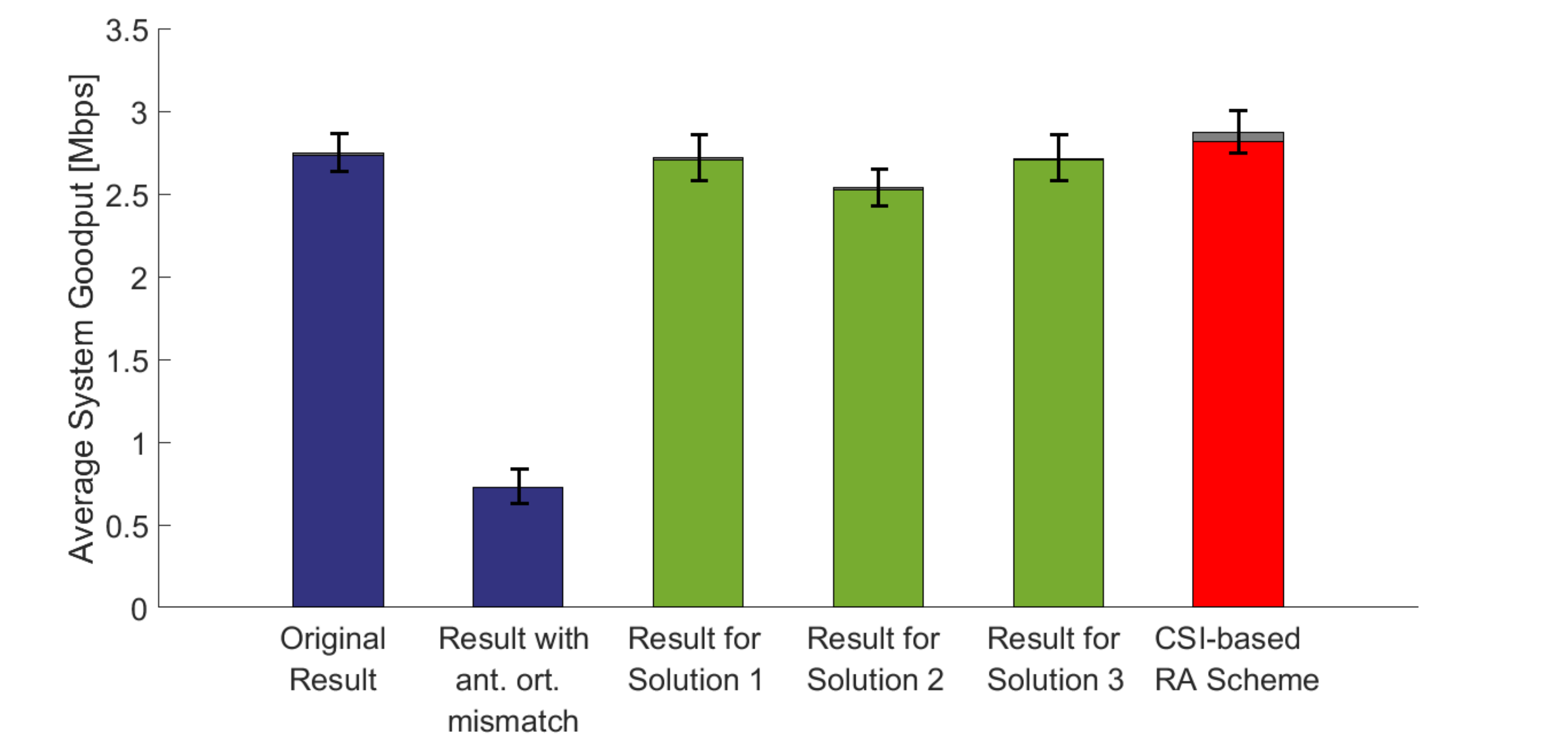}
\caption{Effect of misalignment in UE antenna orientation on the average system goodput, and its mitigation results}
\label{AO1}
\end{figure}

Thus, the antenna orientation of the UE device is expected to affect the system goodput if not known a priori at the serving RRH. 
We therefore study next the impact of such a random orientation of the user antenna on the performance of the learning-based RA scheme.
Figure~\ref{AO1} shows the effect of misalignment in UE antenna orientation information in the training and test datasets for the learning-based RA scheme. 
It can be seen that the average system goodput is adversely affected by the misalignment in antenna orientation at the receiver, with system goodput only being about 27\% of that for the traditional CSI-based scheme for resource allocation. 
This is by the far the biggest impact on the performance of the learning-based RA scheme found in our work. 
Thus, it is important to investigate approaches to mitigate the performance degradation from random antenna orientation at the user.
One way to mitigate the effect of the misalignment in antenna orientation is to train the classification model for the learning-based approach using random UE antenna orientation information, and then test it for data set with random UE antenna orientation information embedded within.
This case is shown as `Solution 1' in Figure~\ref{AO1}. 
In this case, the random UE antenna orientation helps the classifier learn the correlation between different resources and user-related system parameters effectively, thus resulting in the performance gap of only 6\% from the system goodput for the CSI-based RA scheme.

Another option to mitigate the effect of UE antenna orientation is to apply a rotation matrix to adjust the predicted receive filter settings according to realistic UE antenna orientation. 
The mathematical analysis for applying this solution is based on the derivation above. 
The performance result for this method is shown by the bar labelled `Solution 2' in Figure~\ref{AO1}. 
Here, we achieve almost 85\% of the average system goodput compared to the CSI-based scheme, which is fairly good but worse than the performance seen for solution 1. 
A possible reason for this performance loss is the interference present in the system, which makes the solution of only rotating the predicted receive filter for good reception at UE a sub-optimal approach. 
Yet another solution can be applied to mitigate the effect of unknown UE antenna orientation, i.e. by making the UE antenna orientation a part of the input feature vector, exclusively. 
Note that this would require some additional signalling from the user terminal to the AN, for which we do not account for the overhead in the following.
The performance of this approach is shown as `Solution 3' in Figure~\ref{AO1}, where we see that the performance of the proposed technique reaps off almost the same average system goodput as in case of `Solution 1'. 
Since we use the same number of features for random selection in building the decision trees in random forests model, the randomization of trees in the model results in the variation of the obtained system goodput, for the case when UE antenna orientation is embedded or is exclusively incorporated as an input feature for training the random forests model, i.e. for `Solution 1' and `Solution 3', respectively.
We conclude with the remarkable observation that the random antenna orientation can basically deteriorate performance strongly, however, especially by including this effect in the training data, more robust modulation and coding selections can be trained to compensate for this randomness. 
No additional signalling of the user terminal antenna orientation is required.  

\subsection{Change in Channel Statistics}
\begin{figure}[!]
\centering
\includegraphics[width=10cm,height=6cm]{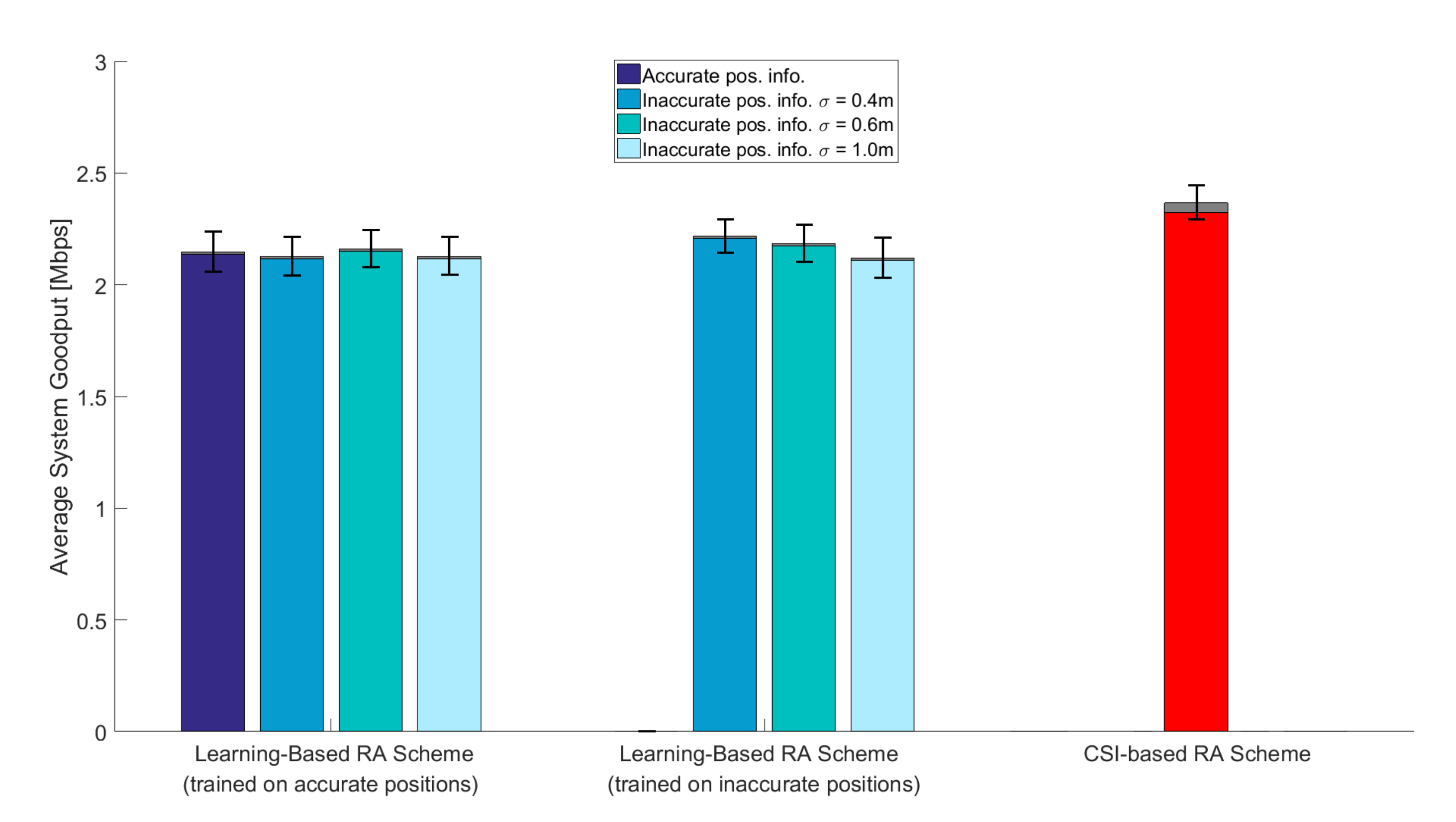}
\caption{Average system goodput for different schemes and various possibilities of available position information for NLOS scenario}
\label{NLOS_result}
\end{figure}

A special case for testing the performance of the proposed scheme is when the LOS links are no more existent between the RRHs and the relevant users in the system. In this case, the specular component is totally neglected when computing the channel matrix for a given RRH-user link, thus resulting in an NLOS scenario. Figure \ref{NLOS_result} shows the average system goodput obtained from the proposed learning-based, as well as the traditional CSI-based RA schemes, for different inaccuracy ranges involved in the acquired user position estimates. We kept the range of user position inaccuracy fairly small in this case, since NLOS consideration is already enough to result in performance degradation using only user-position estimates for resource allocation in the system. Overall, for perfect user position information availability, the proposed scheme still performs fairly well, reaping off almost 90\% of the system goodput obtained using CSI-based RA scheme. The effect on average system performance for different variation in user position inaccuracies, however, does not show a specific trend, because of changing channel statistics in NLOS scenario. But training on inaccurate user position estimates proves to be beneficial in improving the system goodput, in contrast to the effect seen in LOS case, where the learning-based scheme trained only on perfect user position information is enough to guarantee a system performance comparable to the traditional CSI-based RA scheme. 

\section{Conclusions and Future Work}
\label{C&F}
We presented the design of a learning-based RA scheme which has much lower system overhead, as well as lower complexity, than the traditionally used CSI-based RA scheme, because of its dependence on only the acquired user position estimates. Random forests algorithm is used for designing learning-based RA scheme, that works as a self-scheduler for appropriate resource allocation in 5G CRAN system, to serve the different user terminals using only their position information. A comparison analysis was done for the RA scheme based on random forests model and the CSI-based RA scheme, in different contexts. The proposed scheme shows either comparable or significantly better effective system performance compared to the CSI-based RA scheme for different user densities in the system. In terms of the design parameter variations, the proposed scheme is fairly robust to the inaccuracy involved in the user position estimation. Training the random forests model on the data set involving variation in either the system metrics, or the design parameters for the learning-based scheme, is very beneficial in case when the same model trained on fixed system parametrization shows degraded system performance. In general, for LOS or NLOS cases, the proposed scheme is  robust to small error margin involved in the acquired user position information, as well as to the variation in system characterization (such as changing scatterer density). The change in UE antenna orientation affects the performance of the proposed scheme most severely, but the effect can be mitigated by training on top of the UE antenna orientation information, either embedded or provided explicitly, in the training data for constructing the random forest. The performance limitations of the learning-based scheme for extreme channel characterization variation is still an open question, which will be a part of the future work.


\section*{References}

\bibliography{mybibfile}

\appendix
\section{Appendix}
\subsection{Derivation for Rotation Matrix}
\label{App1}

The basic rotation matrices for rotating the vectors by an angle in x-, y- or z-axis, using the right-hand rule, are given as follows \cite{swokowski1979calculus}:

\begin{gather}
\pmb{R}_x (\theta) = \begin{bmatrix}
1 && 0 && 0 \\
0 && cos \theta && -sin \theta\\
0 && sin \theta && cos \theta
\end{bmatrix} ,
\end{gather}

\begin{gather}
\pmb{R}_y (\theta) = \begin{bmatrix}
cos \theta && 0 && sin \theta\\
0 && 1 && 0\\
-sin \theta && 0 && cos \theta
\end{bmatrix} ,
\end{gather}

\begin{gather}
\pmb{R}_z (\theta) = \begin{bmatrix}
cos \theta && -sin \theta && 0\\
sin \theta && cos \theta && 0\\
0 && 0 && 1
\end{bmatrix} .
\end{gather}

For the pre-defined orientation angles of the receive filter for a UE, the rotation matrix can be computed as:

\begin{align}
\pmb{R} = \pmb{R}_z(\phi_0) \times \pmb{R}_y(\theta_0).
\end{align}

Expanding the above expression, we get the following form:

\begin{gather}
\pmb{R} = \begin{bmatrix}
cos \phi_0 && -sin \phi_0 && 0\\
sin \phi_0 && cos \phi_0 && 0\\
0 && 0 && 1
\end{bmatrix} \times
\begin{bmatrix}
cos \theta_0 && 0 && sin \theta_0\\
0 && 1 && 0\\
-sin \theta_0 && 0 && cos \theta_0
\end{bmatrix} 
\end{gather}

\begin{gather}
\pmb{R} = \begin{bmatrix}
cos \phi_0\:cos \theta_0 && -sin \phi_0 && cos \phi_0 \:sin \theta_0\\
sin \phi_0 \:cos \theta_0 && cos \phi_0 && sin \phi_0 \:sin \theta_0\\
-sin \theta_0 && 0 && cos \theta_0
\end{bmatrix}
\end{gather}

Inserting this rotation matrix expression in equation \ref{cos_psi} results in the following expressions for $cos \varphi$ and $sin \varphi$:

\begin{gather}
cos \varphi = \begin{bmatrix}
cos \theta\:cos \phi && cos \theta\: sin \phi && -sin\theta
\end{bmatrix} \times \pmb{R} \times
\begin{bmatrix}
cos \theta'\:cos \phi'\\
cos \theta'\:sin \phi'\\
-sin \theta'
\end{bmatrix},
\end{gather}

\begin{gather}
sin \varphi = \begin{bmatrix}
-sin \phi && cos \phi && 0
\end{bmatrix} \times \pmb{R} \times
\begin{bmatrix}
cos \theta'\:cos \phi'\\
cos \theta'\:sin \phi'\\
-sin \theta'
\end{bmatrix}.
\end{gather}

Simplifying these matrix multiplications gives:

\begin{align}
\begin{split}
cos \varphi = \: & (cos\theta \: cos \theta_0 \: cos\phi'\: +\:sin \theta \:sin \theta_0 ) \times cos \theta'\:cos \phi'\:+\:cos\theta\:sin \phi'\:cos\theta'\:sin\phi'\\
&- \: sin\theta' (cos\theta\:sin\theta_0\:cos\phi'\: - \: sin\theta\:cos\theta_0),
\end{split}
\end{align}

\noindent
and,

\begin{align}
sin \varphi = -cos\theta_0 \: sin \phi'\: cos\theta'\:cos \phi' +\:cos \phi' \:cos \theta'\:sin\phi' \: - \:sin\theta_0 \: sin\phi'\:sin\theta' .
\end{align}

\end{document}